\documentclass[aps,groupaddress,superscriptaddress,reprint,amsmath,amssymb,graphicx,twocolumn,prx]{revtex4-2}
\usepackage{physics}
\usepackage{bm}
\usepackage{bbold}
\usepackage{graphicx}
\usepackage[toc,page]{appendix}
\usepackage{float}
\usepackage{color}

\newcommand{\state}[1]{p_{\boldsymbol{x},\gamma,#1}}

\newcommand{\avg}[1]{\langle #1 \rangle}

\begin{document}
\preprint{APS/123-QED}

\title{Plasticity-induced multistability on fast and slow timescales enables optimal information encoding and spontaneous sequence discrimination}

\author{Giacomo Barzon}
\affiliation{Padova Neuroscience Center, University of Padova, Padova, Italy}

\author{Daniel Maria Busiello}
\thanks{D.M.B. and G.N. contributed equally to this work, and are listed alphabetically}
\affiliation{Max Planck Institute for the Physics of Complex Systems, Dresden, Germany}
\affiliation{Department of Physics and Astronomy ``G. Galilei'', University of Padova, Padova, Italy}

\author{Giorgio Nicoletti}
\thanks{D.M.B. and G.N. contributed equally to this work, and are listed alphabetically}
\affiliation{Quantitative Life Sciences section, The Abdus Salam International Center for Theoretical Physics (ICTP), Trieste, Italy}

\begin{abstract}
\noindent Neural circuits exhibit remarkable computational flexibility, enabling adaptive responses to noisy and ever-changing environmental cues. A fundamental question in neuroscience concerns how a wide range of behaviors can emerge from a relatively limited set of underlying biological mechanisms. In particular, the interaction between activities of neuronal populations and plasticity modulation of synaptic connections may endow neural circuits with a variety of functional responses when coordinated over different characteristic timescales. Here, we develop an information-theoretic framework to quantitatively explore this idea. We consider a stochastic model for neural activities that incorporates the presence of a coupled dynamic plasticity and time-varying stimuli. We show that long-term plasticity modulations play the functional role of steering neural activities towards a regime of optimal information encoding. By constructing the associated phase diagram, we demonstrate that either Hebbian or anti-Hebbian plasticity may become optimal strategies depending on how the external input is projected to the target neural populations. Conversely, short-term plasticity enables the discrimination of temporal ordering in sequences of inputs by navigating the emergent multistable attractor landscape. By allowing a degree of variability in external stimuli, we also highlight the existence of an optimal variability for sequence discrimination at a given plasticity strength. In summary, the timescale of plasticity modulation shapes how inputs are represented in neural activities, thereby fundamentally altering the computational properties of the system. Our approach offers a unifying information-theoretic perspective of the role of plasticity, paving the way for a quantitative understanding of the emergence of complex computations in coupled neuronal-synaptic dynamics.
\end{abstract}

\maketitle

\section{Introduction}
\noindent A fundamental challenge for the brain is to efficiently encode and compute information about the external world in order to guide adaptive behavior. Neuronal populations accomplish this task by generating structured patterns of activity that represent sensory inputs, internal states, and motor plans \cite{olshausen2004sparse, beck2008probabilistic, gallego2017neural, langdon2023unifying}. Rather than acting in isolation, neurons operate within interconnected circuits, and the collective activity of populations emerges through coordinated dynamics that support robust and distributed representations \cite{mante2013context, vyas2020computation, dubreuil2022role, perich2025neural}. These collective operations rely on a network of synaptic connections that fundamentally shape the emergent neural activity. For example, the underlying synaptic interactions determine which activity patterns are expressed, how information is distributed across neurons, and how reliably signals can be decoded downstream \cite{quiroga2007decoding, quian2009extracting, panzeri2022structures}. In this view, understanding neural coding requires considering how the synaptic architecture constrains and enables the formation of population codes within neural circuits.

Crucially, synaptic connectivity is not fixed in time - it changes continuously in response to both internal and external stimuli, providing the fundamental mechanism by which neural circuits flexibly adapt. This activity-dependent modulation of synaptic strength is called plasticity. Plasticity is not a single, uniform process but encompasses multiple, coexisting forms, each driven by distinct biological mechanisms \cite{martin2000synaptic, turrigiano2004homeostatic, holtmaat2009experience, vogels2013inhibitory, froemke2015plasticity, bittner2017behavioral, agnes2024co}. Theoretical and numerical studies have demonstrated that plasticity allows neural populations to explore richer dynamics than static connections alone, leading to novel dynamical phases \cite{laje2013robust, clark2024theory}. Plasticity has also been extensively investigated as a central mechanism through which structured connectivity emerges in neural circuits, shaping their functional properties \cite{ko2013emergence, litwin2014formation, sadeh2021excitatory, lagzi2024emergence}.

In addition to the diversity of plasticity mechanisms, there is another important layer of complexity: the timescales over which they operate \cite{fusi2007neural, zenke2015diverse, spitmaan2020multiple}. For instance, short-term plasticity acts on the order of milliseconds to seconds, dynamically modulating synaptic efficacy in an activity-dependent manner. In contrast, long-term potentiation and depression induce persistent changes that can stabilize learned representations over minutes to hours or even longer. Despite extensive progress in understanding the dynamical effects of plasticity across different timescales, it remains debated how these mechanisms translate into the coding properties of neuronal populations \cite{clopath2010connectivity, soltani2010synaptic, panzeri2010sensory, petersen2013synaptic, gjorgjieva2016computational}. Importantly, since synaptic connectivity shapes the structure of noise correlations, and these correlations in turn constrain the way populations encode information \cite{panzeri2022structures}, plasticity — by modifying synaptic connections — is expected to influence both the amount and the format of information encoded about external signals.

To address these questions, we develop a paradigmatic model of neural populations that incorporates plasticity mechanisms operating at multiple timescales. Using this framework, we show, from an information-theoretic perspective, how plasticity supports distinct computational functions expressed by population activity, depending on the specific timescale at which it operates. Importantly, these computational properties arise from general statistical dependencies in the activity and input distributions, rather than from the specific form of rate dynamics, plasticity rules, or connectivity structure chosen in the model, thus being robust features of a multiscale dynamics.

The remainder of the paper is organized as follows. In Section \ref{sec:main_results}, we briefly outline the main findings of our investigation. Section \ref{sec:model} describes the dynamical equations governing the temporal evolution of the system and presents its analytical solution under different timescale-separation regimes. Sections \ref{sec:slow_results} and \ref{sec:fast_results} focus on the information-theoretic properties of neural activity, first with long-term plasticity (Section \ref{sec:slow_results}) and then with short-term plasticity (Section \ref{sec:fast_results}), highlighting their different computational roles. Finally, Section \ref{sec:discussion} summarizes the results and discusses potential extensions and future perspectives.

\section{Main results} \label{sec:main_results}
\noindent  We study a dynamical model of interacting excitatory and inhibitory neural populations, in which the temporal evolution of population activity is described by stochastic rate equations. The connectivity between populations includes a plastic component, which evolves over time according to specific plasticity rules, allowing populations to adapt their interactions in an activity-dependent manner. Neural populations are also driven by a time-varying external signal, providing a dynamic input that the network must encode. Importantly, each degree of freedom — population activity, plasticity of connections, and external input — evolves on its own characteristic timescale. This hierarchy of timescales gives the system a multiscale character, with fast, intermediate, and slow processes interacting to shape the overall dynamics.

By employing an information-theoretic approach, 
we show that the timescale at which plasticity operates fundamentally alters how inputs are represented in neural activities, thereby shaping the computational properties at the population level. In particular, long-term plasticity modulations play the functional role of steering neural populations toward a global optimum of information encoding. Conversely, short-term plasticity enables efficient discrimination of input sequences by inducing multiple stable attractors in the dynamics, thereby shaping the temporal dimension of information encoding. Remarkably, the emergence of these concurrent functional roles is a sheer consequence of the statistical dependencies stemming from the underlying multiscale nature of the system, independently of the specificity of the implemented dynamics.




\begin{figure*}[t]
    \centering
    \includegraphics[width=1.\textwidth]{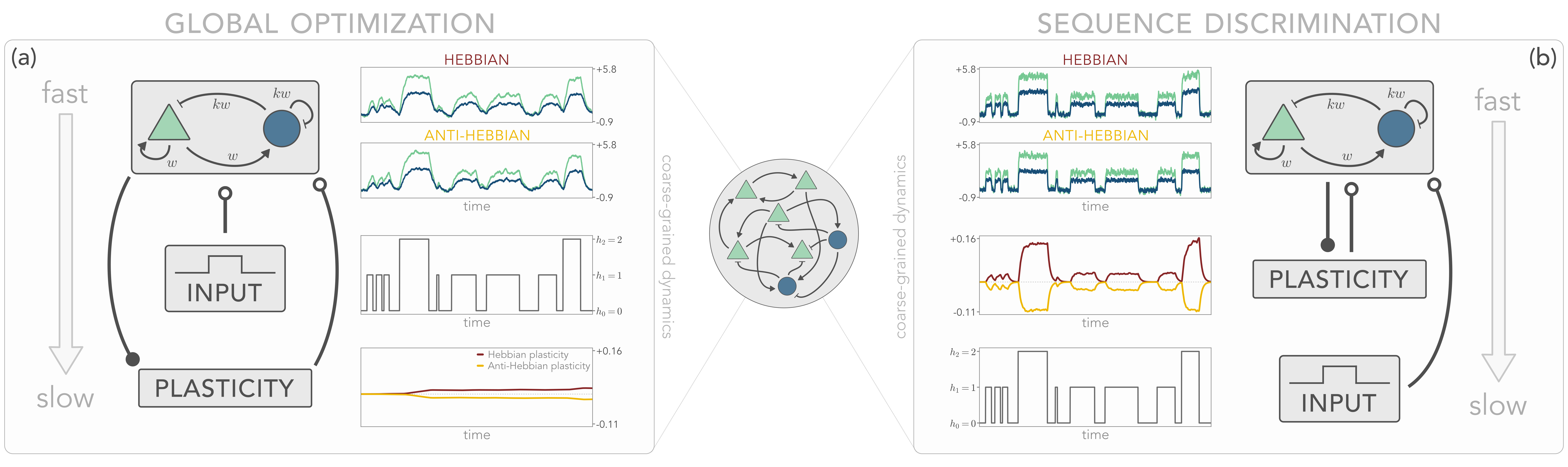}
    \caption{Summary of model and results. The dynamics of a set of coupled excitatory and inhibitory neurons is coarse-grained into the evolution of two coupled subpopulations, Eq.~\eqref{eq:langevin_rho}. To allow information to propagate from input to neural activities, the former always has to evolve more slowly than the latter. (a) Slow plasticity scenario. Plasticity exhibits a slower dynamics than both neural activities, since it acts as a regulatory mechanism, and input. Resulting trajectories of all degrees of freedom are shown for Hebbian and anti-Hebbian cases. In this regime, long-term plasticity modulation helps the system reach a global optimum in terms of information encoding, i.e., the optimal mutual information (see Section \ref{sec:slow_results}). (b) Fast plasticity scenario. In this setting, the input is slower than the plasticity. We show that short-term plasticity modulations are essential to perform sequence discrimination (see Section \ref{sec:fast_results}).}
    \label{fig:summary}
\end{figure*}

\section{Model}
\label{sec:model}
\noindent We consider the activity of two stochastic subpopulations of neurons, one excitatory $x_E$ and one inhibitory $x_I$, coupled by a connectivity matrix $\hat{A}$ and following a Langevin dynamics. Both populations receive a time-varying stimulus $h(t)$, representing the external input that neurons try to encode in their dynamics. We can write the evolution of neural activities in a vectorial form as follows:
\begin{eqnarray}
    \tau \frac{d}{dt} \bm{x} = - \bm{r}\cdot\bm{x} + \hat{A}\,\bm{x} + \sqrt{2 \tau} \,\hat{\sigma} \,\bm{\xi} + h(t) \,\bm{\Lambda} \;,
    \label{eq:langevin_rho}
\end{eqnarray}
where $\bm{x} = \{x_E, x_I\}$, $\bm{r} = \{r_E, r_I\}$ represents the decay of the activity, $\tau$ the characteristic neural timescale, and $\bm{\xi} = \{\xi_E, \xi_I\}$ a Gaussian white noise with amplitude $\hat{\sigma}$ that, for simplicity, we assume to be a diagonal matrix with elements $\sigma_E$ and $\sigma_I$, respectively for excitatory and inhibitory populations. We also introduced the projection vector $\bm{\Lambda} = \{\cos\lambda, \sin\lambda\}$ that controls how the stimulus, potentially stemming from other surrounding brain regions, is received by the neural subpopulations we are describing.

The synaptic connectivity matrix $\hat{A}$ can be parametrized in the following form:
\begin{align}
\label{eq:matrix_A}
    \hat{A} = e^\gamma \begin{pmatrix}
    w & -kw \\
    w & -kw
    \end{pmatrix} \, ,
\end{align}
where $w$ represents the overall excitation strength, $k$ quantifies the relative intensity of the inhibition, and $\gamma$ is a parameter encoding the modulation of coupling strength induced by synaptic plasticity. With these choices, the excitation strength cannot change sign. Such a requirement is biologically motivated by Dale's principle \cite{dale1935pharmacology}. Furthermore, this exponential form for plasticity is similar to exponential gradient descent algorithms, which have been used to optimize biologically-inspired neural networks \cite{kivinen1997exponentiated, pogodin2024synaptic, cornford2024brain}. Although this model represents a simplification of a more complete nonlinear stochastic dynamics \cite{wong2006recurrent, ostojic2011spiking}, it has been studied in several recent works because of its ability to capture essential features of neural connectivity \cite{murphy2009balanced,hennequin2012non, mastrogiuseppe2025stochastic}. Moreover, it has been recently shown that its information-theoretic features strikingly resemble those of a more complex nonlinear scenario in the region of parameters where the system is linearly stable \cite{barzon2025excitation}, which is the regime studied in this manuscript. In particular, the system is linearly stable when $k > k_c(\gamma) = 1 - r/(e^\gamma w)$, so that neural activity is inhibition-dominated. On the other hand,  excitation is too strong for the inhibition to stabilize it for $k < k_c$, with $k = k_c$ corresponding to the point of exact balance between excitation and inhibition.

The synaptic strength $\gamma$ evolves according to the following local plasticity rule:
\begin{equation}
\label{eq:hebbian}
    \tau_p \frac{d}{dt} \gamma = - \gamma + p \, x_E x_I \;,
\end{equation}
where $\tau_p$ is the characteristic synaptic timescale, and $p$ the plasticity strength. This corresponds to a Hebbian plasticity rule when $p>0$, while it is anti-Hebbian for $p<0$ \cite{gerstner2002spiking}. We will show that either one of these rules leads to optimal information encoding in different parameter regimes. Although this is not the only possible choice to model plasticity strength dynamics, our framework can be extended to other local rules. For example, in Appendix \ref{app:oja} we study in more detail the Oja rule, another widely implemented dynamics \cite{oja1982simplified}.

To complete the model description, we need to specify the dynamics of the time-varying input $h(t)$. We model changes in the external environment producing the stimulus as a Markov jump process between a baseline value for the input, $h_0 = 0$, and a set of other $M$ possible values $h_i$. The probability that $h = h_i$ is $\pi_i$ and evolves according to the following Master Equation dynamics:
\begin{equation}
\label{eq:input}
    \frac{d}{dt} \pi_i = \sum_{j=0}^M \left( q_{j\to i} \pi_j - q_{i\to j} \pi_i \right) = \frac{1}{\tau_h} \sum_{j=0}^M \hat{Q}_{ij} \pi_j \;.
\end{equation}
For the sake of simplicity, we assume that $q_{0\to i} = q_\uparrow$ and $q_{i\to 0} = q_\downarrow$ for $i=0, \dots, M$, and all the others are zero. Thus, the timescale associated with the input is given by $\tau_h = (q_\uparrow + q_\downarrow)^{-1}$, and introduce the rescaled transition matrix $\hat{Q}_{ij} = \hat{W}_{ij} / \tau_h$. 


The set of Eqs.~\eqref{eq:langevin_rho}-\eqref{eq:input} can be written in the form of a Fokker-Planck equation describing the evolution of the joint probability distribution $\state{i}$ \cite{nicoletti2021mutual, barzon2025excitation}. This is the probability that, at time $t$, activities take values $\bm{x}$, plasticity $\gamma$, and the external stimulus is equal to $h_i$. It reads~\cite{gardiner1985handbook}:
\begin{align}
\label{eq:FPE}
    \partial_t \state{i} = \left( \frac{\mathcal{L}_{\boldsymbol{x}}(\gamma,h_i)}{\tau} + \frac{\mathcal{L}_{\gamma}(\boldsymbol{x})}{\tau_{p}} + \frac{\mathcal{L}_{i}}{\tau_{h}} \right) \state{h_i}
\end{align}
with
\begin{align}
    &\mathcal{L}_{\bm{x}}(\gamma,h_i) = \bm{\nabla} \cdot \left(\hat{A}\,\bm{x}\right) + \hat{D}\,\nabla^2 \;, \nonumber\\
    &\mathcal{L}_{\gamma}(\bm{x}) = \partial_\gamma \left[\left(-\gamma + p\,x_Ex_I\right) \right] \;, \nonumber\\
    &\mathcal{L}_{i} = \sum_{j=0}^M \hat{Q}_{ij} \;. \nonumber
\end{align}
Here, $\hat{D} = \hat{\sigma} \hat{\sigma}^T$, and $\int d\bm{x} d\gamma \,\state{i} = \pi_{i}$ by construction, so that $\mathcal{L}_i \,\state{i}$ coincides with the right-hand side of Eq.~\eqref{eq:input} once integrated over $\bm{x}$ and $\gamma$. In order to quantify how much information on the external input is encoded in the neural activities distribution, we use the mutual information $I_{\bm{x},h}$ \cite{shannon1948mathematical, blahut1987principles}. This is defined as the Kullback-Leibler divergence between the joint distribution $p_{\bm{x},i} = \int d\gamma \,\state{i}$ and the product of the marginals $p_{\bm{x}}$ and $p_{i}$, respectively obtained by summing $p_{\bm{x},i}$ over $i$ and integrating it over $\bm{x}$:
\begin{equation}
\label{eq:mutual_information}
    I_{\bm{x},h} = \sum_{i=0}^M \int d\bm{x} \, p_{\bm{x},i} \,\log_2 \frac{p_{\bm{x},i}}{p_{\bm{x}} \, p_i}
\end{equation}
where $\log_2$ indicates that we are measuring the mutual information in bits. Recent works have shown that this quantity is a powerful tool for assessing the encoding performance of various biological systems \cite{cheong2011information,mattingly2021escherichia,bauer2023information,nicoletti2024tuning,tkavcik2025information}, particularly in the case of stochastic neural populations \cite{toyoizumi2005generalized, barzon2025excitation}.

Crucially, the behavior of the model fundamentally depends on the different timescales at play. In particular, it can be shown that the input must evolve on a slower timescale than that of neural activities for the neural populations to be able to encode information on the stimulus \cite{barzon2025excitation}. Hence, depending on $\tau_p$, we focus on two limiting scenarios that represent two paradigmatic schemes to understand the role of plasticity modulation in encoding input information: slow and fast plasticity (Figure \ref{fig:summary}). In particular, the slow plasticity scenario, sketched in Figure \ref{fig:summary}a, corresponds to the timescale ordering $\tau_p \gg \tau_h \gg \tau$. When timescales are ordered as $\tau_h \gg \tau_p \gg \tau$, instead, the model represents a dynamics with fast plasticity (Figure \ref{fig:summary}b). We will obtain analytical solutions in both of these limiting cases.


\section{Slow plasticity for global information optimization}
\label{sec:slow_results}
\noindent We first focus on the case of slow plasticity (Figure \ref{fig:summary}a), where the timescales are ordered as $\tau_p \gg \tau_h \gg \tau$. We use a multiple timescale separation approach to solve Eq.~\eqref{eq:FPE}, following a similar procedure to the one introduced in \cite{nicoletti2024information, nicoletti2024gaussian}. The fastest degrees of freedom are excitatory and inhibitory neural activities. Hence, using the slowest timescale $\tau_p$ as a reference, we introduce the expansion parameters $\epsilon_{\bm{x}} = \tau/\tau_p$ and $\epsilon_h = \tau_h/\tau_p$, so that $\epsilon_{\bm{x}} \ll \epsilon_h \ll 1$. Thus, the leading term of Eq.~\eqref{eq:FPE} is of order $\epsilon_{\bm{x}}^{-1}$ and reads:
\begin{equation}
\label{eq:x_steady}
    0 = \mathcal{L}_{\bm{x}}(\gamma,h_i) \,\state{i}^{(0)} \;.
\end{equation}
This induces the following conditional structure in the probability distribution:
\begin{equation}
\label{eq:x_steady_sol}
    \state{i}^{(0)} = p^{\rm st}_{\bm{x}|\gamma, i} \, p_{\gamma, i}^{(0)}
\end{equation}
where the superscript $(\cdot)^{\rm st}$ indicates the steady state with respect to the operator of the non-conditional variable, and $(\cdot)^{(0)}$ recalls that we are computing the solution at the leading order in all expansion parameters. Equation \eqref{eq:x_steady_sol} highlights that neural activities reach a steady-state for every fixed value of both plasticity $\gamma$ and external input $h_i$. Proceeding with the expansion, we have an equation of order $\epsilon_h^{-1}$. By integrating over $\bm{x}$, we obtain that the second fastest degree of freedom is stationary as well:
\begin{equation}
    0 = \sum_{j=0}^M \hat{Q}_{ij} p_{\gamma, i}^{(0)}
\end{equation}
which is solved by
\begin{equation}
    p_{\gamma, i}^{(0)} = \pi^{\rm st}_i \,p_\gamma^{(0)} \;.
\end{equation}
Since the input evolves independently, the operator governing its evolution is the same as in the original equation. Therefore, $\pi_i^{\rm st}$ coincides with the stationary solution of the input dynamics alone. Finally, by integrating on $\bm{x}$, and summing over $i$, at order $1$ we have:
\begin{equation}
\label{eq:dynamic_gamma_slow}
    \partial_{t_p} p^{(0)}_\gamma = \overbrace{\sum_{i=0}^M \left(\int d\bm{x} \,\mathcal{L}_\gamma(\bm{x}) \,p^{\rm st}_{\bm{x}|\gamma,i} \pi^{\rm st}_i\right)}^{\mathcal{L_\gamma^{\rm eff}}} p_\gamma^{(0)} = 0 \;,
\end{equation}
assuming that also the slowest dynamics reaches a stationary state. Here, $t_p = t/\tau_p$ since, in the timescale separation regime, the system solely evolves in the time units of the slowest variable. Notice that the effective operator $\mathcal{L}_\gamma^{\rm eff}$ directly emerges from our approach. The solution to this equation is $p^{(0)}_\gamma = p_\gamma^{\rm eff, st}$ and represents the distribution of the plasticity $\gamma$ when its dynamics feels an infinite trajectory of neural activities, as suggested by $\mathcal{L}^{\rm eff}_\gamma$. The overall solution is:
\begin{equation}
\label{eq:p_slow}
    \state{i}^{\rm slow} = p^{\rm st}_{\bm{x}|\gamma, i} \,\pi^{\rm st}_i \,p^{\rm eff, st}_\gamma
\end{equation}
whose explicit expression is reported in Appendix \ref{app:slowfast_calculations}.


\begin{figure*}[t]
    \centering
    \includegraphics[width=1.\linewidth]{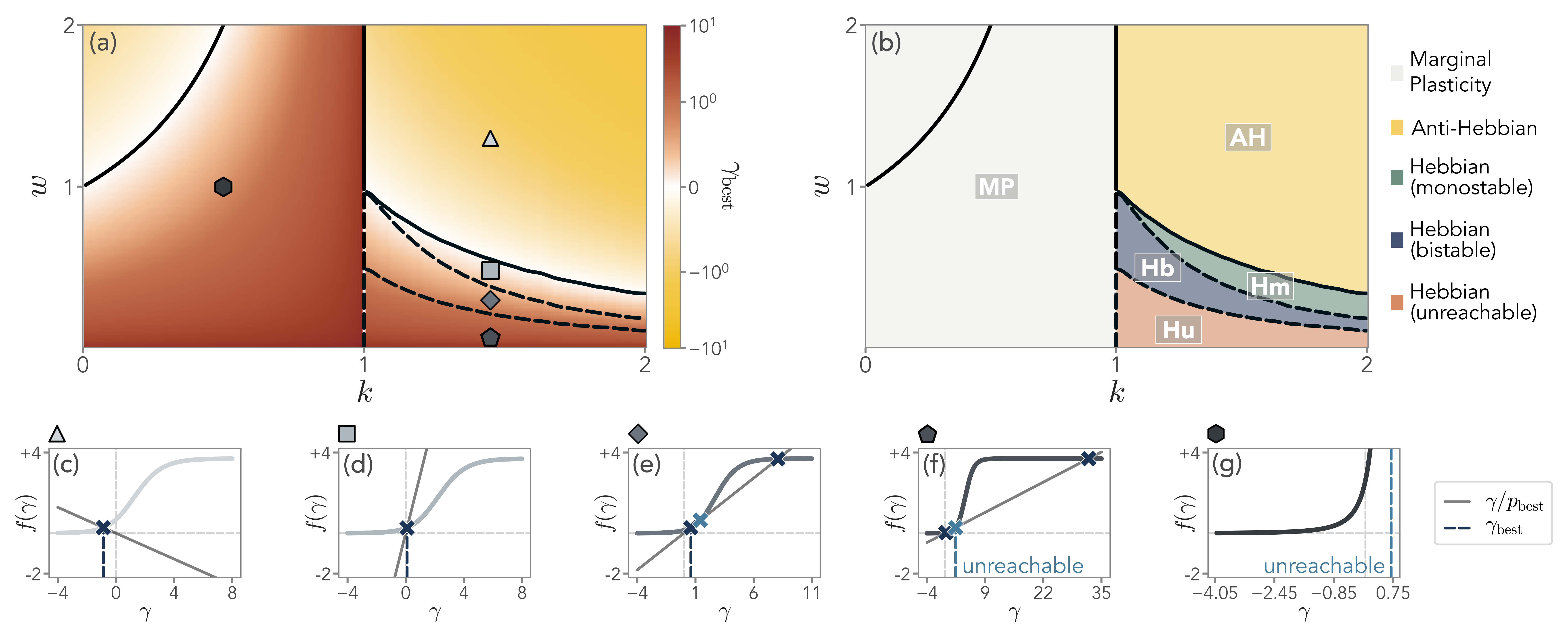}
    \caption{Maximal information by slow plasticity. (a) Contour-plot of the synaptic strength leading to maximal mutual information between input and activities, $\gamma_{\rm best}$. We highlight the existence of different regions (both Hebbian, in red tones, and anti-Hebbian, in yellow tones) and build a corresponding phase diagram in panel (b). (c-g) In each region, we show the graphical solution of the self-consistency equation \eqref{eq:self_consistent}, with $f(\gamma)$ depicted with the same color as the symbol. Notice that, in AH, Hm, and Hb, $\gamma_{\rm best}$ can be stably attained by the system, while in some others, i.e., Hu and MP, it does not constitute a stable solution and, as such, $p_{\rm best}$ is not a feasible optimal plasticity. Here, $r_E = r_I =1$, $\hat{D} = 0.05 \mathbb{1}$, $M=3$, and $\Delta h = 0.5$.}
    \label{fig:p_best}
\end{figure*}

\subsection{Maximal information by Hebbian and anti-Hebbian plasticity}
\noindent Eq.~\eqref{eq:x_steady_sol} shows that neural activities follow a bivariate Gaussian distribution at stationarity,
\begin{equation}
\label{eq:bivariate}
    p_{\bm{x}|\gamma,i}^{\rm st} = \mathcal{N}\left(\bm{\mu}(\gamma, h_i), \hat{\Sigma}(\gamma) \right) \;,
\end{equation}
with a covariance matrix that is independent of the input (see Appendix \ref{app:slowfast_calculations}). The steady-state input distribution $\pi_i^{\rm st}$, following Eq.~\eqref{eq:input}, is:
\begin{equation}
\label{eq:input_st}
    \pi_0^{\rm st} = \frac{q_\downarrow}{q_\downarrow + M q_\uparrow} \qquad \pi_i^{\rm st} = \frac{q_\uparrow}{q_\downarrow + M q_\uparrow} \;
\end{equation}
for $i=1,\dots,M$ different input values. We set $h_i = i \Delta h$ for simplicity. Therefore, the distribution of neural activities given the plasticity value is a Gaussian mixture:
\begin{equation}
\label{eq:mixture_slow}
    p_{\bm{x}|\gamma}^{\rm slow} = \sum_i \pi^{\rm st}_i p_{\bm{x}|\gamma,i} \;.
\end{equation}
Finally, since plasticity follows a deterministic evolution, we have a delta-like stationary distribution for $\gamma$ centered around the second moment of $p_{\bm{x}|\gamma}$ (see Eq.~\eqref{eq:dynamic_gamma_slow}):
\begin{equation}
\label{eq:p_gamma_slow}
    p_\gamma^{\rm eff, st} = \delta\left(\gamma - p\langle x_E x_I \rangle_{\bm{x}|\gamma} \right) \;.
\end{equation}
This is equivalent to a self-consistent equation for $\gamma$:
\begin{equation}
\label{eq:self_consistent}
    \frac{\gamma}{p} = f(\gamma) \; .
\end{equation}
Hence, upon fixing the input values $h_i$ and the neural parameters $w$ and $k$, the plasticity strength $p$ fully determines the properties of the system. The explicit form of $f(\gamma)$ and its properties are reported in Appendix \ref{app:slowfast_calculations}. Crucially, this means that the system can maximize information between inputs and neural activity via long-term plasticity modulations by setting
\begin{equation}
    p_{\rm best} = \gamma_{\rm best}/f(\gamma_{\rm best}) \;,
\end{equation}
where $\gamma_\mathrm{best}$ is the value that maximizes information encoding,
\begin{equation}
    \gamma_{\rm best} = \arg \max_\gamma I_{\bm{x},h}(\gamma) \;.
\end{equation}
In Fig.~\ref{fig:p_best}a, we show $\gamma_{\rm best}$ in the parameter space $(k,w)$, by fixing $\lambda = 0$ for simplicity. Figure \ref{fig:p_best}b presents the associated phase diagram for $p_{\rm best}$. We remark here that ${\rm sign}(\gamma_{\rm best}) = {\rm sign}(p_{\rm best})$ for the selected input projection, so that Hebbian (anti-Hebbian) plasticity coincides with $\gamma_{\rm best}>0$ ($\gamma_{\rm best}<0$). The mutual information has been estimated numerically  by sampling $p^{\rm slow}_{\bm{x}|i}$. Analytical bounds can also be obtained and show similar behaviors with respect to changes in $w$ (see Appendix~\ref{app:bounds}).

When $k>1$, the system is always stable, and different regions can be identified. At large $w$, we have an anti-Hebbian region (AH) in which $p_{\rm best}<0$, i.e., the mutual information is maximized by implementing an anti-Hebbian plasticity rule. In Fig.~\ref{fig:p_best}c, we show the graphical solution of the consistency equation \eqref{eq:self_consistent} for this case. Decreasing $w$, we find a switch to a Hebbian region (solid black line) in which $p_{\rm best}>0$ (Fig.~\ref{fig:p_best}a-b).
Inside this Hebbian region, we can distinguish three different subregions depending on the properties of the solution $\gamma_{\rm best}$. When $\gamma_{\rm best}$ is a stable solution of the consistency equation \eqref{eq:self_consistent}, we have a Hebbian monostable (Hm) phase (green area in Fig.~\ref{fig:p_best}b). At lower $w$, $\gamma_{\rm best}$ becomes one of two possible stable solutions, denoting a Hebbian bistable (Hb) phase (blue area in Fig.~\ref{fig:p_best}b). Finally, at even lower values of $w$, we can identify a Hebbian unreachable (Hu) phase, where $\gamma_{\rm best}$ is unstable for the corresponding $p_{\rm best}$ (red area in Fig.~\ref{fig:p_best}b). Thus, this solution, although corresponding to maximal information encoding, cannot be dynamically reached by the system. We show the shapes of $f(\gamma)$ and the corresponding solutions in Fig.~\ref{fig:p_best}c-f.

When $k<1$, the system is stable only when $k>k_c(\gamma)$ and the mutual information is maximal at the edge of linear stability \cite{barzon2025excitation}. Therefore,
\begin{equation}
    \gamma_{\rm best}\Big|_{k<1} = \log\left( \frac{r}{w (1-k)} \right) \;.
\end{equation}
Formally, this implies that $\gamma_\mathrm{best} > 0$ when $k>k_c$, and viceversa for $k<k_c$.
However, whether positive or negative, $\gamma_{\rm best}$ in this region cannot be stably reached by the system as it coincides with the point at which $f(\gamma_{\rm best}) \to +\infty$, as shown in Fig.~\ref{fig:p_best}g. This divergent behavior is a sheer consequence of the fact that the system is unstable at $\gamma = \gamma_{\rm best}$. Moreover, because of this divergence, $p_{\rm best} \to 0^+$ ($p_{\rm best} \to 0^-$) for $k<1$ under (over) the linear instability line, so that maximal information encoding is formally attained for marginal plasticity (Fig.~\ref{fig:p_best}b). We are not distinguishing between asymptotically small positive and negative $p_{\rm best}$ in Fig.~\ref{fig:p_best}b at this stage. Although this result might sound unexpected, it is tightly related to the fact that $\gamma_{\rm best}$ cannot be reached by the system, and thus the corresponding $p_{\rm best}$ is not a feasible optimal solution.

\begin{figure*}[th!]
    \centering
    \includegraphics[width=1.\textwidth]{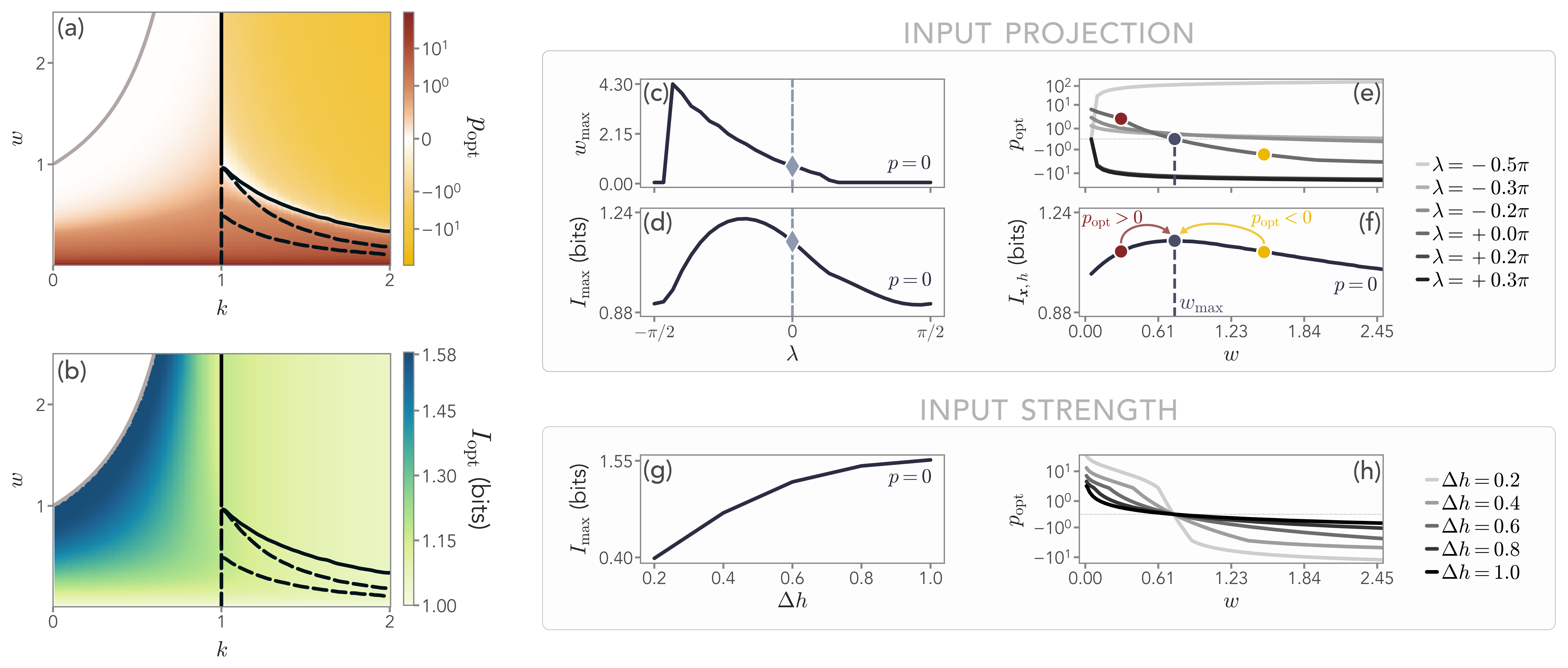}
    \caption{Optimal encoding by slow plasticity and the effect of input features. (a) Contour-plot of the optimal plasticity $p_{\rm opt}$, defined as the one corresponding to a stable $\gamma_{\rm opt}$ with the maximal possible mutual information (Eq.~\eqref{eq:gamma_opt}), as a function of dynamical parameter $k$ and $w$. We see that $p_{\rm opt}$ can be defined even for $k<1$ in the region of linear stability. The corresponding optimal mutual information, $I_{\rm opt}$, is plotted in panel (b). (c) Value of $w$ at which $I_{\bm{x},h}$ is maximal for $p=0$ and $k=1.2$, as a function of the input projection $\lambda$. (d) Maximal value of the mutual information $I_{\rm max}$ for $p=0$ and $k=1.2$, as a function of $\lambda$. We show a peak in encoding performance for a specific direction of projection. (e) We show that the value of $p_{\rm opt}$ may change sign with $\lambda$ (here, $k=1.2$). (f) A positive $p_{\rm opt}$ (red point and arrow) indicates that the corresponding $w<w_{\rm max}$. The opposite holds for a negative $p_{\rm opt}$ (yellow point and arrow). Here, $\lambda = 0$. (g) $I_{\rm max}$ increases with the stimulus spacing $\Delta h$ (here, $p = 0$ and $k = 1.2$). (h) However, the input strength does not affect the nature of the optimal plasticity, whether it is Hebbian or anti-Hebbian. In this figure, we used the same parameters as in Fig.~\ref{fig:p_best} unless specified otherwise.}
    \label{fig:gamma_opt}
\end{figure*}

\subsection{Reachability of optimal encoding}
\noindent Since $\gamma_\mathrm{best}$ may be unreachable for some parameters, we now assume that plasticity attempts to tune the neural populations to a value of $\gamma$ that is both optimal and feasible:
\begin{equation}
\label{eq:gamma_opt}
    \gamma_{\rm opt} = \arg \max_{\gamma \in \mathcal{S}(\gamma)} I_{\bm{x},h}(\gamma)
\end{equation}
where $\mathcal{S}(\gamma)$ denotes the set of $\gamma$ that are stable solutions of Eq.~\eqref{eq:self_consistent} for which $f(\gamma)$ is finite. This provides a value of synaptic strength, $\gamma_{\rm opt}$, that can be dynamically reached by the system and is as close as possible to the one leading to maximal mutual information. Its corresponding value of plasticity strength, defined again by inverting Eq.~\eqref{eq:self_consistent}, is $p_{\rm opt}$. Notice that $p_{\rm opt} = p_{\rm best}$ in the AH, Hm, and Hb regions of Fig.~\ref{fig:p_best}b, since $\gamma_{\rm best}$ is stable and hence is equal to $\gamma_{\rm opt}$. Conversely, $p_{\rm opt} \neq p_{\rm best}$ in the Hu region of Fig.~\ref{fig:p_best}b, as well as for the marginal plasticity phase, since now $\gamma_{\rm opt}$ coincides with the stable solution closest to the unstable one leading to maximal $I_{\bm{x},h}$ (see Fig.~\ref{fig:p_best}f). See Appendix \ref{app:slowfast_calculations} for a graphical representation.

In Figure \ref{fig:gamma_opt}a we plot the optimal plasticity strength as a function of parameter $w$ and $\gamma$. At variance with the previous case, $p_{\rm opt}$ can be consistently defined in the MP region ($k<1$) as long as the system remains linearly stable. Following the definition in Eq.~\eqref{eq:gamma_opt}, $\gamma_{\rm opt}$ is the point at which the stable and the unstable solutions collapse, thus $p_{\rm opt}$ is obtained accordingly from Eq.~\eqref{eq:self_consistent} (see Appendix \ref{app:slowfast_calculations}). Notice that, in the linearly unstable region (above the gray line in Fig.~\ref{fig:gamma_opt}a), every $p\to 0^-$ strictly different from zero is formally stable and reachable. As a consequence, optimality is achieved by marginal plasticity, and therefore, it is not possible to formally define a $p_\mathrm{opt}$ in this region. In Fig.~\ref{fig:gamma_opt}b, we show the value of the mutual information at the optimal plasticity, $I_{\rm opt} \equiv I_{\bm{x},h}|_{p\to p_{\rm opt}}$. Crucially, we find that the optimal mutual information increases as the system approaches excitation-inhibition balance at $k = k_c (\gamma_\mathrm{opt})$.

\subsection{The effect of input projection and strength}
\noindent Input features modify the encoding performance of neural activities. In our framework, there are two main parameters to control the input: its projection $\lambda$, determining how it is relayed to excitatory and inhibitory populations from other brain regions; and $\Delta h$, modulating the strength of different inputs as $h_i = i \Delta h$.

We first focus on the role of $\lambda$ in the absence of plasticity, i.e., $p=0$. Figure \ref{fig:gamma_opt}c shows the value of $w = w_{\rm max}$ at which the information $I_{\bm{x},h}$ attains its maximum value $I_{\rm max}$ for $p=0$ at fixed $k$. We find that $I_{\rm max}$ shows a peak at a specific value of $\lambda$, suggesting the existence of a preferential direction $\bm{\Lambda}$ associated with optimal coding features (see Fig.~\ref{fig:gamma_opt}d). After introducing plasticity, we find that 
the region in which the optimal plasticity is Hebbian (anti-Hebbian), i.e., $p_{\rm opt}>0$ ($p_{\rm opt}<0$), depends on the input projection (see Fig.~\ref{fig:gamma_opt}e). This is due to the fact that, at fixed $k$, the value of $w_\mathrm{max}(\lambda)$ can be either lower or higher than $w$ depending on $\lambda$, thus leading to a positive or negative $p_\mathrm{opt}$ (see Fig.~\ref{fig:gamma_opt}f).

The effect of input strength on $I_{\rm max}$ in the absence of plasticity is shown in Fig.~\ref{fig:gamma_opt}g. As expected, the mutual information increases with the spacing between different stimuli. In Fig.~\ref{fig:gamma_opt}h, we study $p_{\rm opt}$ as a function of $w$, i.e., at fixed $k$ for different values of $\Delta h$. Interestingly, the input strength does not modify the nature of the optimal plasticity, e.g., $p_{\rm opt}$ does not change sign with $\Delta h$. Hence, both Hebbian and anti-Hebbian regions are preserved independently of the magnitude of the stimuli that the system receives.

\section{Fast plasticity for sequence discrimination}
\label{sec:fast_results}
\noindent To model fast plasticity, we now derive an explicit solution to Eq.~\eqref{eq:FPE} when timescales are ordered as $\tau_h \gg \tau_p \gg \tau$ (Figure \ref{fig:summary}b). Using the same procedure as for the slow plasticity case, we introduce the expansion parameters $\varepsilon_{\bm x} = \tau/\tau_h$ and $\varepsilon_p = \tau_p/\tau_h$, so that $\varepsilon_{\bm x}\ll \varepsilon_{p} \ll 1$. The leading order equation is of order $\varepsilon_{\bm{x}}^{-1}$ and, once again, coincides with Eq.~\eqref{eq:x_steady}. This sets the distribution of the fastest variable $\bm{x}$ at stationarity at each fixed value of plasticity $\gamma$ and input $h_i$, so that the resulting conditional structure is the same as in Eq.~\eqref{eq:x_steady_sol}.

Since now the plasticity is associated with the intermediate timescale, by proceeding with the expansion, we arrive at an equation of order $\varepsilon_p^{-1}$. By integrating over the fast variable $\bm{x}$, we obtain:
\begin{equation}
\label{eq:gamma_eff_i}
    0 = \overbrace{\left( \int d\bm{x} \mathcal{L}_\gamma(\bm{x}) \,p^{\rm st}_{\bm x|\gamma, i}\right)}^{\mathcal{L}_{\gamma|i}^{\rm eff}} p^{(0)}_{\gamma,i} \;.
\end{equation}
As before, this effective operator takes into account the effect of neural activities by averaging over an infinite trajectory. However, since the input is the slowest variable at play, $\mathcal{L}^{\rm eff}_{\gamma|i}$ has to be determined for every fixed value $h_i$. As a consequence, $\gamma$ reaches a different stationary distribution for each value of the input. Indeed,
\begin{equation}
    p^{(0)}_{\gamma, i} = p^{\rm eff, st}_{\gamma|i} p^{(0)}_i
\end{equation}
solves Eq.~\eqref{eq:gamma_eff_i}. Finally, $p^{(0)}_i$ is determined by the solution of the input dynamics, which is independent of all the other degrees of freedom. Assuming that it also reaches a steady state, we have
\begin{equation}
    \partial_{t_h} p^{(0)}_i = \sum_{j=0}^M Q_{ij} p^{(0)}_j = 0 \; ,
\end{equation}
where $t_h = t/\tau_h$, so that $p^{(0)}_i = \pi^{\rm st}_i$. Thus, the overall solution in this case reads:
\begin{equation}
\label{eq:p_fast}
    p^{\rm fast}_{\bm{x},\gamma,i} = p_{\bm x|\gamma, i}^{\rm st} \,p^{\rm eff, st}_{\gamma|i} \,\pi^{\rm st}_i \;,
\end{equation}
where all the terms are specified in Appendix \ref{app:slowfast_calculations} \footnote{As a remark, in what follows we will use the superscripts $\cdot^{\rm eff, st}$ and $\cdot^{\rm st}$ only for the probability whose conditional structure stems directly from the timescale separation approach. For all the other probability distributions obtained by marginalization, we use the superscript $\cdot^{\rm slow}$ or $\cdot^{\rm fast}$ depending on the plasticity timescale.}.


\begin{figure*}[t]
    \centering
    \includegraphics[width=1.\textwidth]{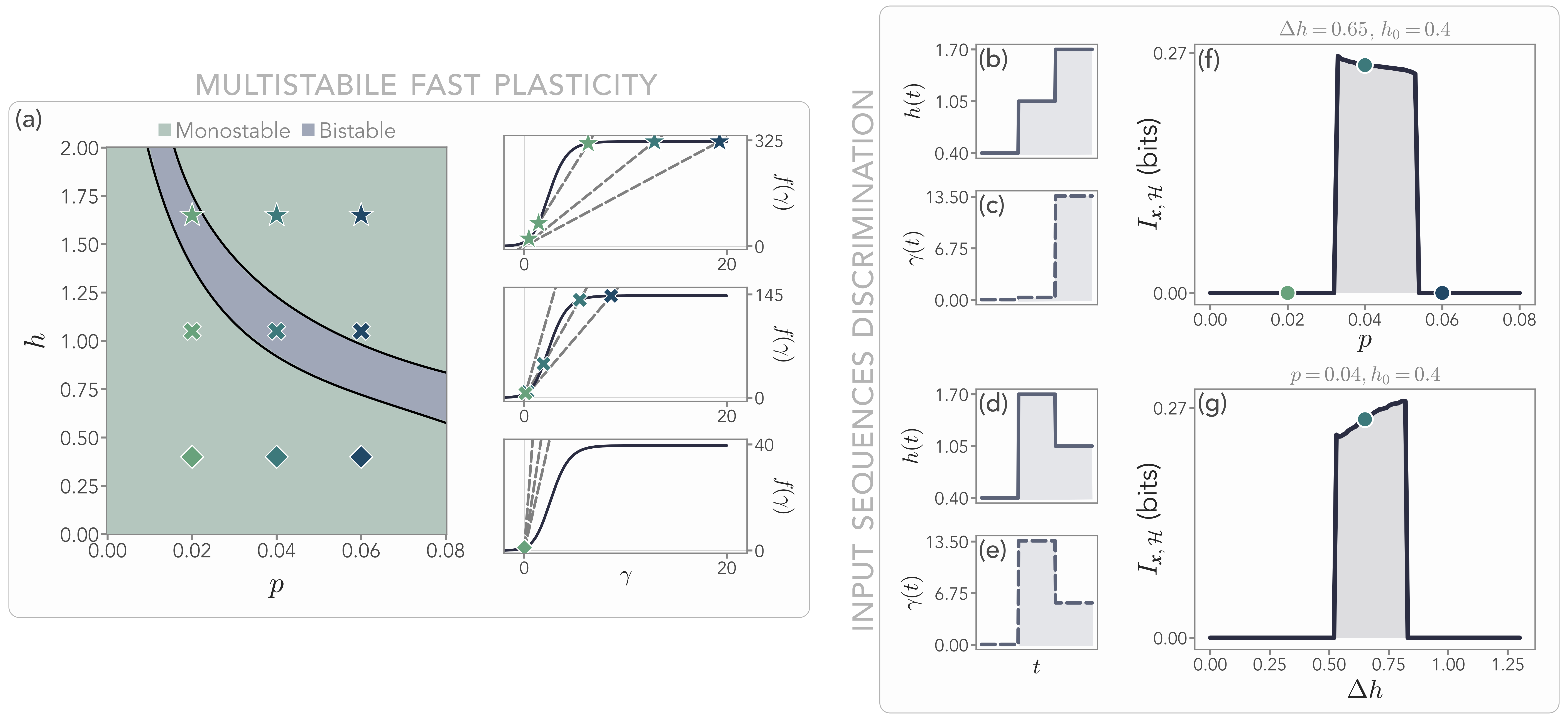}
    \caption{Fast plasticity allows for sequence discrimination through multistability. (a) Three sets of inputs are shown for different $p$ and a fixed $\Delta h = 0.65$. In blue, we highlight the multistability region. Subpanels show the graphical solution to the self-consistency equation (Eq.~\eqref{eq:self_consistency_i}) for all the inputs. (b) The set of inputs of panel (a) is presented as a sequence in ascending order for $p = 0.04$. (c) The corresponding sequences of synaptic strengths. (d) The same set of inputs as in (b) but in a different order. (e) The sequence of associated synaptic strengths changes as well, due to the presence of multistability. (f) Mutual information between sequences of neural activities and sequences of inputs, $I_{\bm{x},\mathcal{H}}$, as a function of $p$ for $\Delta h$ fixed as in (a). Colored dots indicate the specific sets shown in (a) with the same colors. The mutual information is non-zero only when the multistability region can be effectively exploited. (g) $I_{\bm{x},\mathcal{H}}$ as a function of $\Delta h$ for $p=0.04$. Also in this case, the presence of multistability induces a non-zero mutual information. Colored dots indicate the value $\Delta h = 0.65$ as shown in (a). In this figure, $D = 0.1 \mathbb{1}$, $k=1.1$, $w = 2$, $r_E = r_I = 1$.}
    \label{fig:fast_plasticity}
\end{figure*}

\subsection{Multistability enables encoding of sequence information}
\noindent The structure of the probability distribution is now given by Eq.~\eqref{eq:p_fast}. Neural activities follow, once again, the bivariate Gaussian distribution of Eq.~\eqref{eq:bivariate} for any given value of $\gamma$ and $h_i$. Since the plasticity evolves on a faster timescale than the input, the probability of finding a synaptic plasticity $\gamma$, given $h = h_i$, follows the delta-like distribution:
\begin{equation}
\label{eq:p_delta_fast}
    p^{\rm eff, st}_{\gamma|i} = \delta\left(\gamma - p\langle x_E x_I \rangle_{\bm{x}|\gamma, i} \right) \;.
\end{equation}
This corresponds to $M$ self-consistency equations, one for each input values, 
\begin{equation}
\label{eq:self_consistency_i}
    \gamma/p = f_i(\gamma) \;.
\end{equation}
that determine $M$ different $\gamma_i$ for every value of plasticity $p$ \footnote{Notice that $\langle f_i(\gamma)\rangle_i = f(\gamma)$ (see Eq.~\eqref{eq:self_consistent}).}. Depending on both $p$ and the input strength $h$, the system might now exhibit multistability for some inputs $i$, leading to multiple solutions for $\gamma_i$ (see Fig.~\ref{fig:fast_plasticity}a). Finally, the stationary input distribution remains independent of any other variable and equals the one in Eq.~\eqref{eq:input_st}.

One crucial difference with respect to the slow plasticity scenario is that now the system explicitly depends on the sequence of external stimuli. To fix the ideas, consider sequences composed of a fixed number of $L$ inputs $h^{(\mu)}$, for $\mu = 1, \dots, L$, each presented for an interval $\tau_\mu$ such that the total duration is fixed, i.e., $\sum_{\mu=1}^L \tau_\mu = T$. We also assume that $h^{(\mu)}$ is extracted from the same pool of $M$ stimuli, so that $h^{(\mu)} \in \{h_0, \dots, h_M\}$ with $h_i = i \Delta h$ as before. These inputs induce a corresponding sequence in the plasticity $\Gamma = \{\gamma^{(1)}, \dots, \gamma^{(L)}\}$. Due to multistability, in Fig.~\ref{fig:fast_plasticity}b-e, we show that the same set of inputs presented in different orders will lead to different temporal sequences of the plasticity.

This is in contrast to the slow plasticity case, where $p^{\rm eff, st}_{\gamma}$ only depended on an integrated input. The first consequence is that the presence of short-term plasticity undermines the notion of global optimization explored for long-term modulation. However, a fundamental advantage is that the statistics of neural populations now directly reflects the specific input sequence at hand. Thus, we can quantify the information-theoretic performance in discriminating different sequences extracted from the same set of inputs. To this end, we compute
\begin{equation}
\label{eq:pfast_sequence}
    p^{\rm fast}_{\bm{x},\Gamma,\mathcal{H}} = p^{\rm st}_{\bm{x}|\Gamma,\mathcal{H}} \,p^{\rm eff,st}_{\Gamma|\mathcal{H}} \,p^{\rm st}_{\mathcal{H}} \;,
\end{equation}
which is the probability of simultaneously observing neural activities $\bm{x}$, a certain sequence of synaptic strengths $\Gamma$, and a certain sequence of inputs $\mathcal{H}$. In particular, 
\begin{equation}
\label{eq:p_fast_xh}
    p^{\rm fast}_{\bm{x}|\mathcal{H}} = \sum_{h_i\in\mathcal{H}} \frac{\tau_i}{T} p_{\bm{x}|i}^{\rm fast} \;,
\end{equation}
where $p^{\rm st}_{\bm{x}|i}$ is the probability of observing the activity $\bm{x}$ given an input $h_i$, as defined in Eq.~\eqref{eq:p_fast}. This simple solution stems from the fact that neural activities relax instantaneously for a given value of the input during the sequence due to the timescale separation regime. Finally, we assume that sequences are distributed uniformly:
\begin{equation}
    p^{\rm st}_{\mathcal{H}} = \frac{1}{\dim\left(\mathcal{H}_T^{(L)}\right)} \;,
\end{equation}
where $\mathcal{H}_T^{(L)}$ is the space of all possible sequences of a given length and duration. Notice that, while each term in Eq.~\eqref{eq:pfast_sequence} is different from those in Eq.~\eqref{eq:p_fast}, the probabilistic structure depends solely on the timescale ordering and, as such, is conserved. 

\begin{figure*}[ht!]
    \centering\includegraphics[width=1.\textwidth]{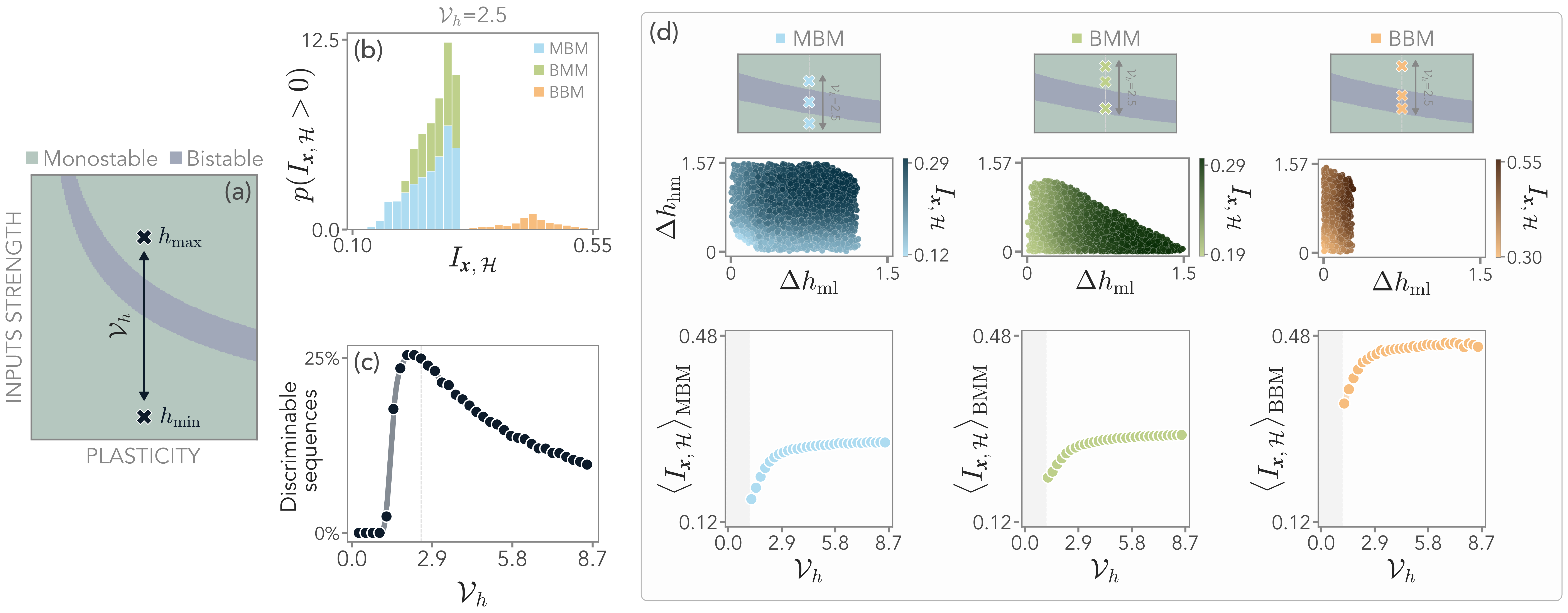}
    \caption{Effect of input variability on sequence discrimination. (a) Monostable and bistable regions in the space of plasticity VS input strength. At a fixed $h$, the corresponding plasticity $p$ might exhibit bistability, i.e., two stable solutions for $\gamma$. Input variability $\mathcal{V}_h$ at a fixed $p$ is defined as the range in which sequences are drawn. (b) Distribution of non-zero value of the mutual information between a sequence of activities $\bm{x}$ and a given sequence of inputs $\mathcal{H}$. $I_{\bm{x},\mathcal{H}}>0$ signals the ability of the system to discriminate temporal ordering of $\mathcal{H}$. Only three types of sequences allow for order discrimination - MBM, BMM, and BBM, determined by which of the inputs lie in the bistability region (see panel (d)). These three classes identify three different clusters. (c) Number of sequences for which discrimination works. Increasing the variability, it becomes more likely to extract sequences with all inputs away from the bistability region and, as such, non-discriminable. The vertical gray line indicates $\mathcal{V}_h = 2.5$. (d) An example of an MBM, BMM, and BBM sequence. Below, the mutual information within each cluster as a function of input spacings $\Delta h_{hm} = h_{\rm high} - h_{\rm med}$ and $\Delta h_{\rm med} - h_{\rm low}$. For BMM sequences, only $\Delta h_{ml}$ regulates $I_{\bm{x},\mathcal{H}}$ since this is the only spacing in which multistability is involved. Lower panels show that the average mutual information within each cluster saturates by increasing the input variability.}
    \label{fig:mutual_sequence}
\end{figure*}

Therefore, we can quantify how much information is shared between $\bm{x}$ and $\mathcal{H}$ as:
\begin{equation}
    I_{\bm{x},\mathcal{H}} = \sum_{\mathcal{H}\in\mathcal{H}_T^{(L)}} \int d\bm{x} \,p^{\rm fast}_{\bm{x},\mathcal{H}} \log_2 \frac{p^{\rm fast}_{\bm{x},\mathcal{H}}}{p^{\rm fast}_{\bm{x}} \,p^{\rm st}_{\mathcal{H}}}
\end{equation}
where $p^{\rm fast}_{\bm{x},\mathcal{H}} = p^{\rm fast}_{\bm{x}|\mathcal{H}} p^{\rm st}_{\mathcal{H}}$ by definition. $I_{\bm{x},\mathcal{H}}$ can be estimated numerically as before (see Appendix \ref{app:bounds}). When $I_{\bm{x},\mathcal{H}}=0$, all sequences in the considered set are equivalent, i.e., neural activities cannot reflect any feature related to input ordering.
Conversely, a non-zero value of this mutual information indicates that the uniform distribution of input sequences induces a non-uniform distribution of synaptic strength sequences due to the interplay between input ordering and multistability, which in turn signals the ability of the system to encode information on $\mathcal{H}$.

In Fig.~\ref{fig:fast_plasticity}f, we show $I_{\bm{x},\mathcal{H}}$ at different values of $p$, for sequences with $L=3$ and inputs extracted from the set of values in Fig.~\ref{fig:fast_plasticity}a without repetitions. We also initialize the system at the steady-state of $h=0$. Remarkably, the mutual information is non-zero only when one of the inputs falls within the multistability region. Notice that this does not hold for the case $p=0.02$ -  only the largest input is multistable, and its corresponding largest $\gamma$ cannot be explored since the system has been initialized at zero stimulus. Figure \ref{fig:fast_plasticity}g shows how the mutual information changes as a function of $\Delta h$ at a fixed $p=0.04$. Once again, we notice that, when multistability can be effectively exploited, the statistics of neural activities carries information on sequences of presented inputs.

\subsection{Effect of input variability}
\noindent Finally, we study how the intrinsic variability of external stimuli affects the capability ot the system to discriminate different sequences by means of short-term plasticity. In particular, we consider input sequences that follow a prescribed distribution within a maximum and a minimum magnitude, respectively $h_{\rm max}$ and $h_{\rm min}$. We define the input variability as $\mathcal{V}_h = h_{\rm max} - h_{\rm min}$ (see Fig.~\ref{fig:mutual_sequence}a).

Variability may affect information. To illustrate our ideas, we construct a sequence by drawing $L=3$ inputs - $h_{\rm low}$, $h_{\rm med}$, and $h_{\rm high}$ - from a uniform distribution in the range $\mathcal{V}_h = h_{\rm max}$, with $h_{\rm min} = 0$.
The ensemble of possible sequences $\mathcal{H}_T^{(L)}$ for these inputs is constructed, as before, by considering all possible orderings of the selected values. If we compute $I_{\bm{x},\mathcal{H}}$ for each randomly extracted sequence, we obtain a distribution of mutual information that solely depends on $h_\mathrm{max}$ and $h_\mathrm{min}$. In Fig.~\ref{fig:mutual_sequence}b, we show that, at fixed $\mathcal{V}_h$, we can identify three different clusters of sequences leading to non-zero mutual information, i.e., those for which order discrimination is effective. These clusters differ from each other by the position of the input leading to a multistable $\gamma$: for BMM sequences, this is the lower input; for MBM, the intermediate one; for BBM, both the lower and the intermediate inputs lie in the multistable region (see Fig.~\ref{fig:mutual_sequence}c). The mutual information within each cluster increases as a function of the distances between the inputs, $\Delta h_{\rm hm} = h_{\rm high} - h_{\rm med}$ and $\Delta h_{\rm ml} = h_{\rm med} - h_{\rm low}$. In particular, it depends on the distance involving the input leading to an unstable $\gamma$. For this reason, the mutual information in the cluster BMM only depends on $\Delta h_{\rm ml}$ (see Fig.~\ref{fig:mutual_sequence}c). As a second observation, the average value of the mutual information within each cluster - $\langle I_{\bm{x},\mathcal{H}}\rangle_{\rm BMM}$,  $\langle I_{\bm{x},\mathcal{H}}\rangle_{\rm MBM}$, and  $\langle I_{\bm{x},\mathcal{H}}\rangle_{\rm BBM}$ - increases with $\mathcal{V}_h$ (see Fig.~\ref{fig:mutual_sequence}c).

Furthermore, there is an additional effect when increasing input variability. Since the multistability region exists for a finite range of input values, at fixed $p$, the larger the intervals from which stimuli are drawn, the larger the probability that they will all end up away from the multistable region. Since multistability is essential to allow for sequence discrimination, a larger $\mathcal{V}_h$ will necessarily correspond to a larger number of sequences with zero mutual information. At the same time, smaller variability will not allow input to be drawn from the multistable region. The combined outcome of these two competing effects translates into the existence of a peak in the number of sequences for which discrimination works ($I_{\bm{x},\mathcal{H}}>0$), as shown in Fig.~\ref{fig:mutual_sequence}d. This result hints at the fact that, given the internal dynamical parameters and the plasticity strength, there is a variability allowing for optimal encoding of sequence information.

\section{Discussion}
\label{sec:discussion}
\noindent A central challenge in neuroscience is to understand how neural circuits generate diverse and flexible computations from a limited set of underlying biological mechanisms. Neuronal activity and synaptic processes do not act on a single characteristic timescale but instead unfold over milliseconds, seconds, or even hours and days. This hierarchy of timescales profoundly shapes circuit dynamics, as fast processes like synaptic transmission interact with slower mechanisms such as homeostatic regulation and long-term synaptic modifications. Because these processes are inherently interdependent, their combination can endow the same circuit with very different functional repertoires depending on how the timescales are coordinated \cite{panzeri2010sensory, gjorgjieva2016computational}. Explicitly accounting for this multiscale structure is therefore essential for understanding how neural populations encode, integrate, and transform information.

Building on this perspective, we developed a flexible framework to investigate, from an information-theoretic standpoint, how synaptic plasticity mechanisms operating at different timescales shape the functional and computational roles of neural populations. Our results reveal that different computational properties arise from the statistical dependencies across timescales, which are shaped by the specific hierarchy of the relevant degrees of freedom governing the system, i.e., neural activity, external input, and plasticity modulation in our setting~\cite{nicoletti2024information, barzon2025excitation}.

With long-term plasticity, the synaptic changes depend primarily on the average input, effectively acting as a global modulation of synaptic strength. We showed that such modulation can steer the system toward regimes of optimal information encoding. Biologically, these two levers may correspond to different regulatory mechanisms: the plasticity parameter itself could be controlled by neuromodulatory substances, which gate or modulate the overall expression of plasticity \cite{fremaux2016neuromodulated, pedrosa2017role, brzosko2019neuromodulation}, while the tuning of input strength and projection could be implemented through top-down feedback signals that instruct synaptic changes in a context-dependent manner \cite{lillicrap2016random, whittington2017approximation, miconi2017biologically, borra2024task}.

In contrast, short-term plasticity gives rise to a completely different phenomenology. In this regime, plasticity rapidly adapts to each value of the external input, effectively becoming quenched to the stimulus and endowing the system with a repertoire of multistable attractors. This multistability enables sequence discrimination, since different input orders drive distinct trajectories across the attractor landscape. Importantly, sequence discrimination can occur on the fly, because the attractor structure is dynamically shaped by ongoing dynamics rather than being rigidly embedded in the recurrent connectivity \cite{gillett2020characteristics, recanatesi2022metastable}. These dynamics resonate with recent theories of working memory, which propose that short-term information storage in recurrent networks relies on fast forms of synaptic plasticity rather than exclusively on persistent activity \cite{mongillo2008synaptic, barak2014working, lansner2023fast}. Our findings thus align with the synaptic theory of working memory while providing a mechanistic account of how rapid plasticity can support flexible temporal coding. They are also consistent with previous numerical investigations of temporal coding and sequence processing \cite{buonomano2000decoding, gutig2006tempotron, buonomano2009state, ballintyn2019spatiotemporal}, thereby offering a unifying theoretical framework that grounds those results in a principled, information-theoretic perspective.



An intriguing future direction is to extend our framework to jointly study fast and slow plasticity, possibly even with different functional forms. Combining plasticity mechanisms operating on multiple timescales has been shown to improve performance compared to classical single-timescale strategies, particularly in terms of robustness and generalization when learning new target signals \cite{bicknell2025fast}. Moreover, the coexistence of fast and slow plasticity has been demonstrated to be sufficient for assembly formation and memory recall in balanced networks \cite{zenke2015diverse}. Our framework provides a natural setting to investigate how these complementary processes interact to support flexible computations.

Additionally, in our current formulation, plasticity is described by a single global parameter that uniformly scales all the connections between neural populations. An immediate extension would be to introduce distinct plasticity dynamics for specific cell types \cite{eckmann2024synapse}. For example, targeting inhibitory populations could be particularly relevant, as they play a central role in balancing stability and flexibility in cortical circuits \cite{vogels2013inhibitory}. Taking this idea further, assigning plasticity mechanisms at the level of individual connections would naturally expand the dimensionality of the plasticity space, allowing for a broader range of possible network configurations and potentially providing a richer substrate for flexible and efficient information processing.

\appendix

\renewcommand{\thefigure}{A\arabic{figure}}
\setcounter{figure}{0}

\widetext

\section{Solution of Fokker-Planck equation under timescale separation}
\label{app:slowfast_calculations}

\noindent To solve the full dynamical model described in Eq.~\eqref{eq:FPE}, we follow a multiple timescale separation approach \cite{ nicoletti2024information}. We recall that in the slow plasticity scenario, the timescales are ordered as $\tau_p \gg \tau_h \gg \tau$, while they are $\tau_h \gg \tau_p \gg \tau$ in the fast plasticity limit. First, we have to solve for the fastest degrees of freedom, that is, the neural activity $\boldsymbol{x}$ in both scenarios. This leads the bivariate Gaussian distribution $p^{\text{st}}_{\boldsymbol{x}|\gamma,i}$ in Eq.~\eqref{eq:bivariate} with mean $\boldsymbol{\mu}(\gamma,h_i)=(\mu_E(\gamma,h_i),\mu_I(\gamma,h_i))$:
\begin{align}
    \begin{cases}
        \mu_E(\gamma,h_i) = \dfrac{h_i}{r \beta_+} [(r+kw e^{\gamma}) \cos \lambda - kw e^{\gamma} \sin \lambda]\\
        \mu_I(\gamma,h_i) = \dfrac{h_i}{r \beta_+} [ w e^{\gamma} \cos \lambda + (r- w e^{\gamma})\sin \lambda]
    \end{cases}
    \label{eq:neuronal_averages_reduced}
\end{align}
and covariance $\hat\Sigma(\gamma)$:
\begin{align}
    \hat\Sigma(\gamma) &= \dfrac{D}{r \beta_+ (2r + w_+)} \begin{pmatrix}
        2r^2 + wr e^\gamma (3k-1) + 2k^2w^2e^{2\gamma} & w e^{\gamma}(r - kr + 2kwe^{\gamma}) \\
        w e^{\gamma} (r -kr + 2kwe^{\gamma}) & 2r^2 + wr e^{\gamma}(k-3) + 2w^2e^{2\gamma}
    \end{pmatrix}
    \label{eq:sigma_stat}
\end{align}
where $\beta_+=r+w_{+}$ and $w_{+}=(k-1)e^{\gamma}w$. In what follows, we use $\hat{\Sigma}_{A B} = \langle x_A x_B \rangle_{\bm{x}|\gamma}$ to indicate matrix elements.

Then, we have to solve for the other degrees of freedom. In the case of slow plasticity (see Sec.~\ref{sec:slow_results}), the plasticity is described in Eq.~\eqref{eq:p_gamma_slow} by a delta-like stationary distribution for $\gamma$ centered around the second moment of the mixture Gaussian $p_{\bm{x}|\gamma}$ in Eq.~\eqref{eq:mixture_slow}, therefore $\gamma$ satisfies the self-consistency equation:
\begin{align}
\label{eq:slow_sc}
    \frac{\gamma}{p} := f(\gamma) &= \hat{\Sigma}_{EI}(\gamma) + \sum_i \pi_i \mu_E(\gamma,h_i) \mu_I(\gamma,h_i) \\ \nonumber
    &= \dfrac{D w e^{\gamma} (r -kr + 2kwe^{\gamma})}{r \beta_+ (2r + w_+)} + \frac{\langle h^2 \rangle }{r^2 \beta_+^2}   \biggl[r \cos \lambda + e^\gamma k w(\cos \lambda- \sin \lambda) \biggl]  \biggl[e^\gamma w (\cos \lambda - \sin \lambda) + r \sin \lambda) \biggl]
\end{align}
where $\langle h^2 \rangle = \sum_i \pi_i h^2_{i}$. We notice that $f(\gamma)$ is defined for all $\gamma$ such that the whole system is linearly stable. Thus, for $k > 1$ it is defined for all $\gamma \in \mathbb{R}$, whereas for $k < 1$ it is defined only for $\gamma \in \left( -\infty, \gamma_c \right)$ where $\gamma_c=\log\left( \frac{r}{w (1-k)} \right)$. Moreover, it is interesting to observe that the limits for small $\gamma$ and large $\gamma$ (when they exist) are:
\begin{align}
\label{eq:bounds_f_gamma}
    \begin{cases}
        \lim\limits_{\gamma\to-\infty} f(\gamma) = \dfrac{\cos \lambda \sin \lambda}{r^2} \langle h^2 \rangle \\
        \lim\limits_{\gamma\to+\infty} f(\gamma) =  k\dfrac{(1 - \sin 2\lambda)\langle h^2 \rangle +2Dr}{r^2(k-1)^2}
    \end{cases}
\end{align}

Instead, in the case of fast plasticity (see Sec.~\ref{sec:fast_results}), the probability of finding a synaptic plasticity $\gamma$, given $h = h_i$, follows similarly a delta-like stationary distribution shown in Eq.~\eqref{eq:p_delta_fast}, that, for each input values $h_i$, is solved by the self-consistency equation:
\begin{align}
\label{eq:fast_sc}
    \frac{\gamma}{p} := f_i(\gamma)
    &= \hat{\Sigma}_{EI}(\gamma) + \mu_E(\gamma,h_i) \mu_I(\gamma,h_i) \\ \nonumber
    &= \dfrac{D w e^{\gamma} (r -kr + 2kwe^{\gamma})}{r \beta_+ (2r + w_+)} + \frac{h_i^2}{r^2 \beta_+^2}   \biggl[r \cos \lambda + e^\gamma k w(\cos \lambda- \sin \lambda) \biggl]  \biggl[e^\gamma w (\cos \lambda - \sin \lambda) + r \sin \lambda) \biggl]
\end{align}

\begin{figure}
    \centering
    \includegraphics[width=0.6\linewidth]{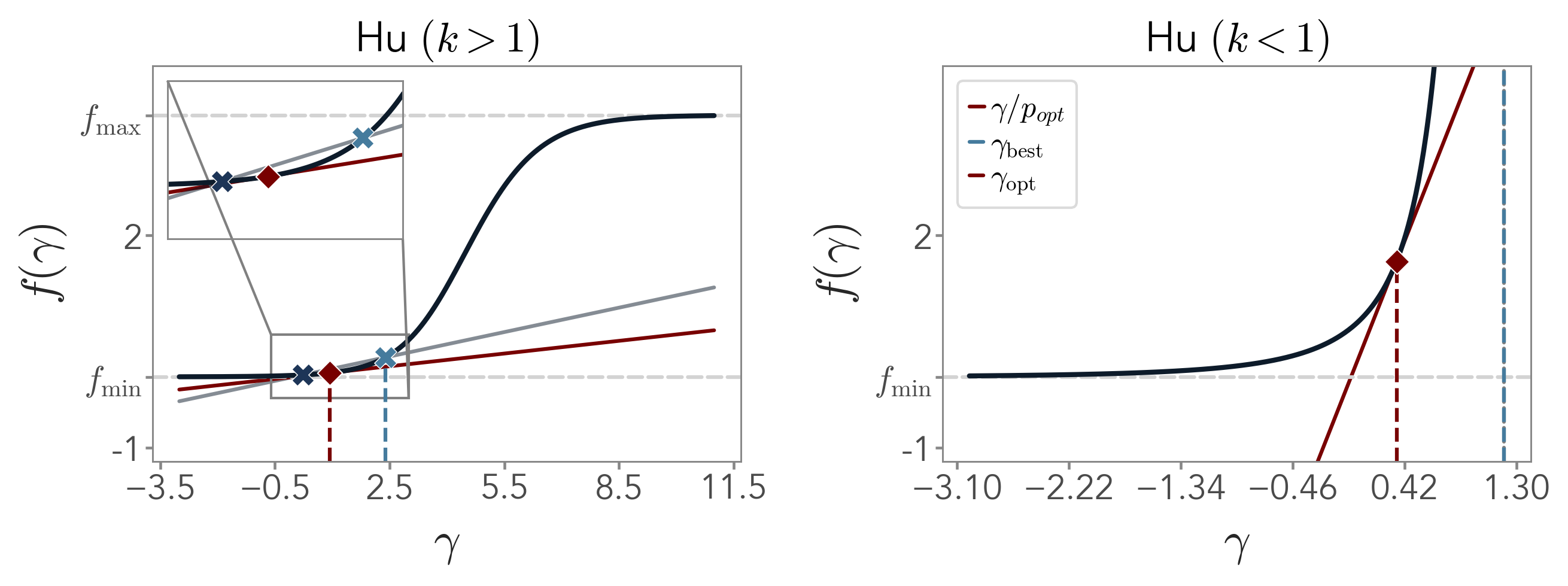}
    \caption{Graphical solution of the self-consistency equation Eq.~\eqref{eq:slow_sc}in the slow-plasticity regime. For a given plasticity parameter $p$, admissible solutions correspond to the intersections between $f(\gamma)$ and the line through the origin with slope $1/p$. In the inhibition-stabilized regime ($k > 1$), the system can admit either one stable solution or three solutions (two stable and one unstable). For $k < 1$, $f(\gamma)$ is defined only up to $\gamma_c$, where it diverges, and the system can admit either one stable solution or two solutions (one stable and one unstable). We highlight cases in which the information-maximal $\gamma_{\text{best}}$ corresponds to an unstable solution and is therefore not dynamically accessible. In such cases, the effective optimal value $\gamma_{\text{opt}}$ is the closest stable solution to $\gamma_{\text{best}}$, which can be estimated from the tangent condition between $f(\gamma)$ and the line of slope $1/p$. Dashed gray lines indicate the analytical bounds derived in Eq.~\eqref{eq:bounds_f_gamma}.
    }
    \label{fig:app_consistency}
\end{figure}

\section{Numerical computation and analytical bounds of mutual information}
\label{app:bounds}

\noindent With slow plasticity, the mutual information Eq.~\eqref{eq:mutual_information} can be recast as:
\begin{align}
    I_{\bm{x},h} &= \sum_{i=0}^M \int d\bm{x} \, p^{\rm slow}_{\bm{x}|i}  \,\pi^{\rm st}_i \,\log_2 \frac{p^{\rm slow}_{\bm{x}|i}}{p^{\rm slow}_{\bm{x}} } \\
    &= \mathbb{E}_{\bm{x},i} \biggl[ \log_2 \frac{p^{\rm slow}_{\bm{x}|i}}{p^{\rm slow}_{\bm{x}} } \biggl]
\end{align}
where $p^{\rm slow}_{x|i}=\int d \gamma \,p^{\rm slow}_{\bm{x}|\gamma,i}$. In general, Eq.~\eqref{eq:mutual_information} does not admit an analytical expression. Therefore, we numerically compute it from the statistics of $p^{\rm slow}_{\bm{x}|i}$ using importance sampling \cite{mackay1998introduction}. However, we can analytically estimate lower and upper bounds starting from the bounds on the entropy of Gaussian mixtures proposed in~\cite{kolchinsky2017estimating}, in particular by estimating the Chernoff-$\alpha$ divergence and the Kullback-Leibler divergence between the mixture components.

\begin{figure}
    \centering
    \includegraphics[width=0.6\linewidth]{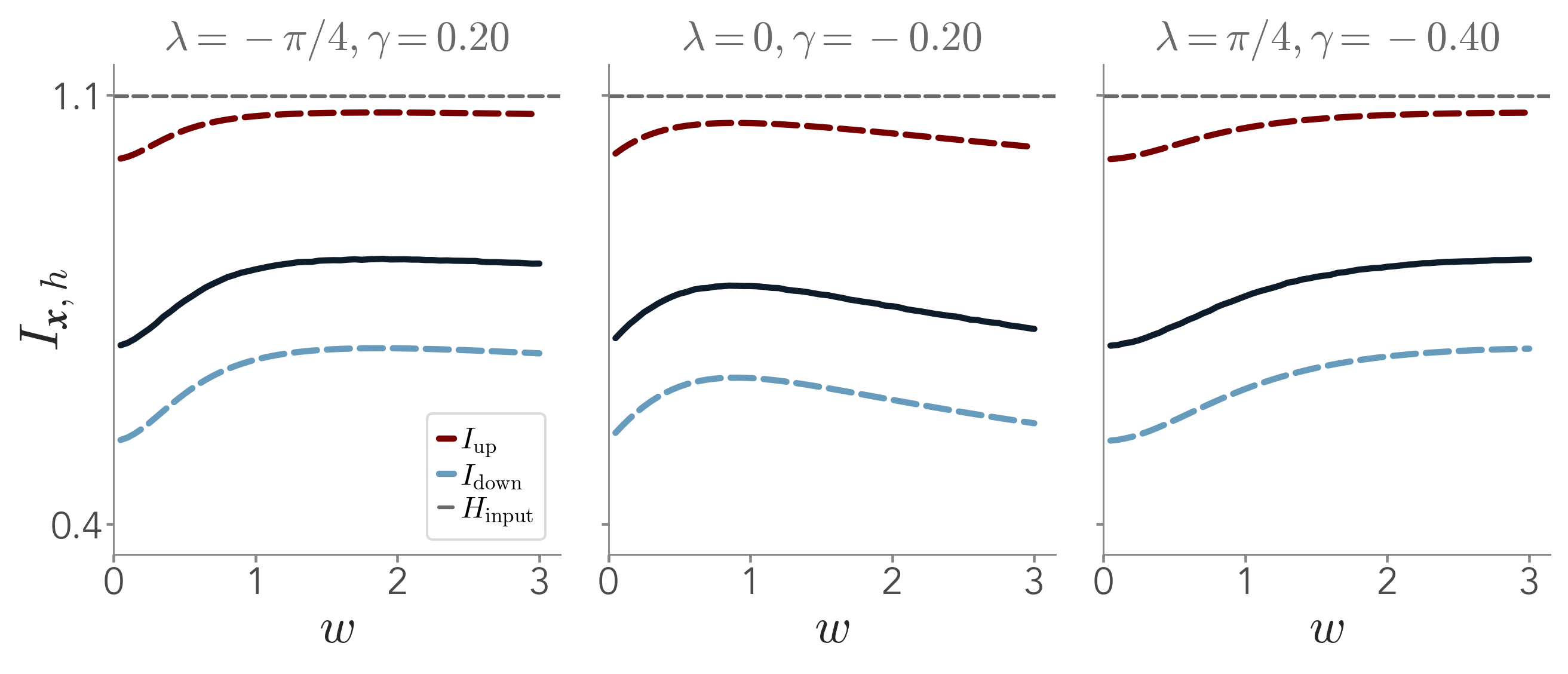}
    \caption{Mutual information in the slow-plasticity regime. Mutual information $I_{\bm{x},h}$, estimated numerically with importance sampling (solid lines), is shown together with the analytical bound Eq.~\eqref{eq:bounds_general} (dashed lines) as a function of $w$. In all cases, the mutual information follows the same trend as the bound and remains robust across different input projections $\lambda$ and plasticity parameter $\gamma$, here illustrated for $k = 1.2$.}
    \label{fig:app_mutual}
\end{figure}

Indeed, the mutual information is lower-bounded by:
\begin{align}
    I_{\boldsymbol{x},h} \geq - \sum_i \pi_i \,\log \biggl[ \sum_j \pi_j \,e^{- C_{0.5}(p^{\rm slow}_{\boldsymbol{x}|i}||p^{\rm slow}_{\boldsymbol{x}|j})} \biggl] = I_\mathrm{low} \ ,
\end{align}
where:
\begin{align}
    C_\alpha(p,q) = - \int dx p^\alpha(x) q^{1-\alpha}(x)
\end{align}
is the Chernoff-$\alpha$ divergence, with $\alpha \in [0,1]$, and the stricter bound for Gaussian mixtures is for $\alpha=0.5$ (which is also called Bhattacharyya distance \cite{bhattacharyya1943measure}). Instead, an upper bound is:
\begin{align}
    I_{\boldsymbol{x},h} \leq - \sum_i \pi_i \,\log \biggl[ \sum_j \pi_j \,e^{- D_\mathrm{KL}(p^{\rm slow}_{\boldsymbol{x}|i}||p^{\rm slow}_{\boldsymbol{x}|j})} \biggl] = I_\mathrm{up} \; .
\end{align}
Each component $p^{\rm slow}_{\boldsymbol{x}|i}$ shares the same covariance matrix $\hat{\Sigma}(\gamma)$. For multivariate Gaussian with different means $\boldsymbol{\mu}_i$ and $\boldsymbol{\mu}_j$ and same covariance $\hat\Sigma$, the Chernoff-$\alpha$ divergence reduces to:
\begin{align}
    C_\alpha(\mathcal{N}(\boldsymbol{\mu}_i, \hat\Sigma), \mathcal{N}(\boldsymbol{\mu}_j, \hat\Sigma)) = \dfrac{\alpha(1-\alpha)}{2} (\boldsymbol{\mu}_i - \boldsymbol{\mu}_j)^T\hat\Sigma^{-1} (\boldsymbol{\mu}_i - \boldsymbol{\mu}_j) \ .
\end{align}
Similarly, the Kullback-Leibler divergence reduces to:
\begin{align}
    D_\mathrm{KL}(\mathcal{N}(\boldsymbol{\mu}_i, \hat\Sigma), \mathcal{N}(\boldsymbol{\mu}_j, \hat\Sigma)) = \dfrac{1}{2} (\boldsymbol{\mu}_i - \boldsymbol{\mu}_j)^T\hat\Sigma^{-1} (\boldsymbol{\mu}_i - \boldsymbol{\mu}_j)
    \label{eq:kl_divergence}
\end{align}
which, except for a prefactor, is identical to the Chernoff-$0.5$ divergence. Thus, by inserting the explicit expressions for the stationary mean and covariance of the mixture components in the slow-input limit, we obtain:
\begin{equation}
    I_\mathrm{low} = I^{(b)}(\eta/4), \qquad I_\mathrm{up} = I^{(b)}(\eta)
    \label{eq:bounds_general}
\end{equation}
where:
\begin{equation}
    I^{(b)}(\eta) = - \sum_{i=0}^M \pi_i^\mathrm{st} \log \biggl[ \sum_{j=0}^M \pi_j^\mathrm{st} e^{-(j-i)^2 \eta} \biggl] \ ,
    \label{eq:bounds_specific}
\end{equation}
and:
\begin{align}
    \eta = \frac{\Delta h^2}{4Dr} \dfrac{2r+w_+}{\beta_+} \dfrac{2r^2+rw e^{\gamma} [(k+1)\cos 2\lambda - (k-1)(\sin 2\lambda-2)] -e^{2\gamma} w^2 (k^2+1)(2 \sin 2\lambda -1)
    }{2r^2+2(k-1)rw e^\gamma+(k^2+1)w^2 e^{2\gamma}} \ .
\end{align}

Similarly, in the fast plasticity scenario, the mutual information Eq.~\eqref{eq:mutual_information} can be recast as:
\begin{align}
    I_{\bm{x},\mathcal{H}} &= \sum_{\mathcal{H}\in\mathcal{H}_T^{(L)}} \int d\bm{x} \,p^{\rm fast}_{\bm{x}|\mathcal{H}} \, p^{\rm st}_{\mathcal{H}} \, \log_2 \frac{p^{\rm fast}_{\bm{x}|\mathcal{H}}}{p^{\rm fast}_{\bm{x}}} \\
    &= \mathbb{E}_{\bm{x},\mathcal{H}} \biggl[ \log_2 \frac{p^{\rm fast}_{\bm{x}|\mathcal{H}}}{p^{\rm fast}_{\bm{x}}} \biggl]
\end{align}
where $p^{\rm fast}_{\bm{x}|\mathcal{H}}$ is itself a mixture Gaussian over the step in the sequence $\mathcal{H}$ as shown in Eq.\eqref{eq:p_fast_xh}.

\begin{figure}
    \centering
    \includegraphics[width=0.6\linewidth]{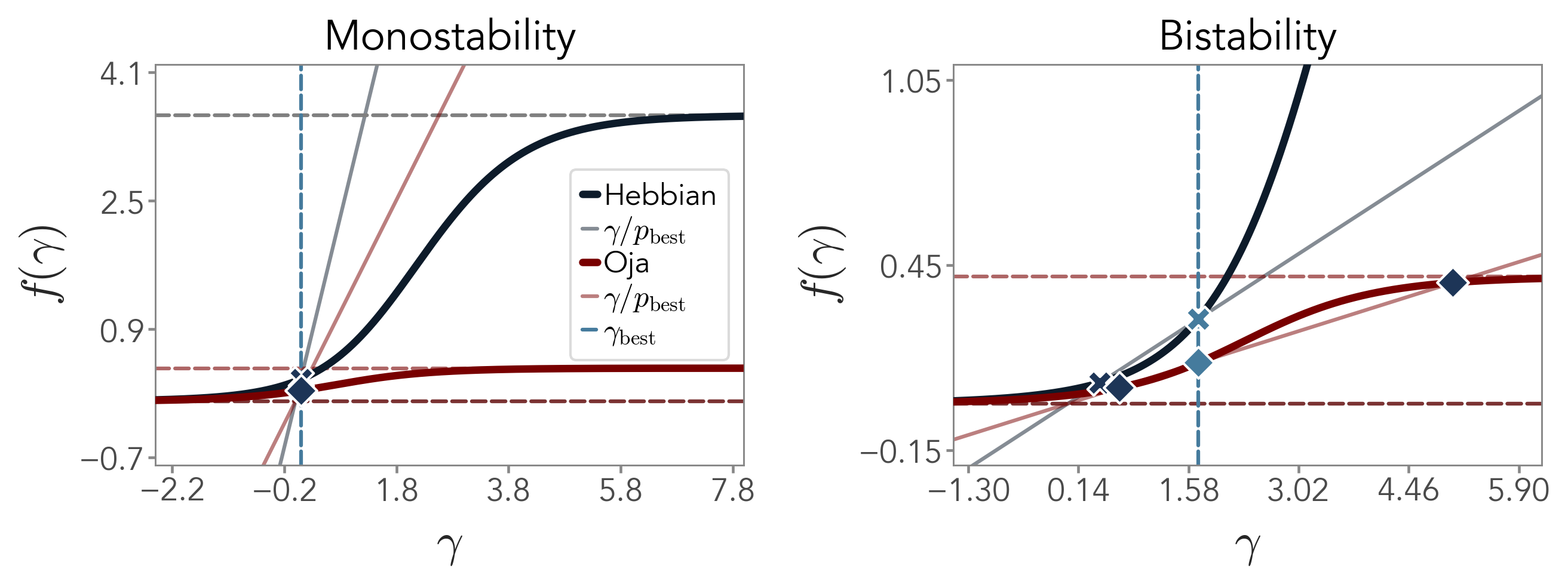}
    \caption{Comparison between Hebbian and Oja plasticity. Graphical solutions of the self-consistency equation for $\gamma$ under Hebbian and Oja plasticity. While Oja plasticity yields a more restricted $f(\gamma)$ (see Eq.~\eqref{eq:oja_slow_sc}, dashed red bounds are shown Eq.~\eqref{eq:bounds_f_gamma_oja}) compared to the Hebbian case (dashed gray bounds), the qualitative behavior remains similar. Since $\gamma_{\text{best}}$ is independent of the specific plasticity rule, what changes is the value of $p$ at which it can be reached. Both mono-stable and bi-stable regimes are illustrated, showing that the overall phenomenology is preserved across the two plasticity forms.}
    \label{fig:app_oja}
\end{figure}

\section{Alternative synaptic plasticity rule}
\label{app:oja}

\noindent Given the plethora of plasticity mechanisms reported in the literature, our model can be readily adapted to different functional forms beyond the Hebbian rule defined in Eq.~\eqref{eq:hebbian}. As an example, we consider a multiplicative regularization term in the plasticity dynamics, inspired by Oja’s normalization \cite{oja1982simplified}:
\begin{align}
    \partial_t \gamma = - (1 + x_E^2 + x_I^2) \gamma + p x_E x_I
\end{align}
Focusing on the slow-plasticity regime, the plasticity is now described by:
\begin{equation}
    p_\gamma^{\rm eff, st} = \delta\left( (1+\avg{x_E^2}_{\boldsymbol{x|\gamma}}+\avg{x_I^2}_{\boldsymbol{x|\gamma}})\gamma - p\langle x_E x_I \rangle_{\bm{x}|\gamma} \right) \;.
\end{equation}
which is solved by the self-consistency equation:
\begin{align}
\label{eq:oja_slow_sc}
    \frac{\gamma}{p} &= \frac{\hat{\Sigma}_{EI}(\gamma) + \sum_i \pi_i \mu_E(\gamma,h_i) \mu_I(\gamma,h_i)}{1 + \hat{\Sigma}_{EE}(\gamma) + \hat{\Sigma}_{II}(\gamma) + \sum_i \pi_i [\mu_E^2(\gamma,h_i) + \mu_I^2(\gamma,h_i) ]} \ ,
\end{align}
where $\mu_A$ and $\hat{\Sigma}_{AB}$ are the ones defined above from the mean Eq.~\eqref{eq:neuronal_averages_reduced} and the covariance matrix Eq.~\eqref{eq:sigma_stat}. Equation \eqref{eq:oja_slow_sc} is bounded by:
\begin{align}
\label{eq:bounds_f_gamma_oja}
    \begin{cases}
        \lim\limits_{\gamma\to-\infty} f(\gamma) = \dfrac{\cos \lambda \sin \lambda}{r^2 +2Dr+ \langle h^2 \rangle} \langle h^2 \rangle \\
        \lim\limits_{\gamma\to+\infty} f(\gamma) =  k\dfrac{2Dr + (1 - \sin 2\lambda)\langle h^2 \rangle}{(k-1)^2r^2 + (k^2+1) [2Dr + (1-\sin 2\lambda)\avg{h^2}]}
    \end{cases} 
\end{align}
Similarly, in the case of fast plasticity (see Sec.~\ref{sec:fast_results}), $\gamma$ is described by:
\begin{equation}
    p^{\rm eff, st}_{\gamma|i} = \delta\left( (1+\avg{x_E^2}_{\boldsymbol{x}|\gamma,i}+\avg{x_I^2}_{\boldsymbol{x}|\gamma,i})\gamma - p\langle x_E x_I \rangle_{\bm{x}|\gamma, i} \right) \;,
\end{equation}
and the self-consistency equation becomes:
\begin{align}
    \frac{\gamma}{p} &= \frac{\hat{\Sigma}_{EI}(\gamma) + \mu_E(\gamma,h_i) \mu_I(\gamma,h_i)}{1 + \hat{\Sigma}_{EE}(\gamma) + \hat{\Sigma}_{II}(\gamma) + \mu_E^2(\gamma,h_i) + \mu_I^2(\gamma,h_i)} \;.
\end{align}


%

\end{document}